\documentclass[aps,prl,twocolumn,showpacs,groupedaddress]{revtex4} 
\usepackage{graphicx}  % needed for figures
\usepackage{subfigure}
\usepackage{dcolumn}   % needed for some tables
\usepackage{bm}        % for math
\usepackage{amssymb}   % for math

\newcommand{\ststbar}{\ensuremath{\tilde{t}_{1}\bar{\tilde{t}}_{1}}}
\newcommand{\stopone}{\ensuremath{\tilde{t}_{1}}}
\newcommand{\chargino}{\ensuremath{\tilde{\chi}^{\pm}_{1}}}
\newcommand{\charginoplus}{\ensuremath{\tilde{\chi}^{+}_{1}}}
\newcommand{\neutralino}{\ensuremath{\tilde{\chi}^{0}_{1}}}
\newcommand{\ttbar}{\ensuremath{t\bar{t}}}
\newcommand{\ppbar}{\ensuremath{p\bar{p}}}
\newcommand{\pt}{\ensuremath{p_{T}}}
\newcommand{\met}{\ensuremath{\slash\kern-.7emE_{T}}}
\newcommand{\mujets}{$\mu+$jets}
\newcommand{\ejets}{$e+$jets}

\newcommand{\pythia}{\textsc{pythia}}
\newcommand{\alpgen}{\textsc{alpgen}}
\newcommand{\comphep}{\textsc{CompHEP-Singletop}}
\newcommand{\geant}{\textsc{geant}}

\begin{document}

\hspace{5.2in} \mbox{Fermilab-Pub-09/005-E}

\title{Search for admixture of scalar top quarks\\ 
in the {\boldmath $t\bar{t}$} lepton+jets final state at 
{\boldmath $\sqrt{s}=1.96$}~TeV}
% LIST_OF_AUTHORS_R2.TEX                 11/25/08           
%
\author{V.M.~Abazov$^{36}$}
\author{B.~Abbott$^{75}$}
\author{M.~Abolins$^{65}$}
\author{B.S.~Acharya$^{29}$}
\author{M.~Adams$^{51}$}
\author{T.~Adams$^{49}$}
\author{E.~Aguilo$^{6}$}
\author{M.~Ahsan$^{59}$}
\author{G.D.~Alexeev$^{36}$}
\author{G.~Alkhazov$^{40}$}
% ...

\affiliation{\vspace{0.1 in}(The D\O\ Collaboration)\vspace{0.1 in}}
% ...
\affiliation{$^{6}$University of Alberta, Edmonton, Alberta, Canada,
                Simon Fraser University, Burnaby, British Columbia, Canada,
                York University, Toronto, Ontario, Canada, and
                McGill University, Montreal, Quebec, Canada}
% ...
\affiliation{$^{29}$Tata Institute of Fundamental Research, Mumbai, India}
% ...
\affiliation{$^{36}$Joint Institute for Nuclear Research, Dubna, Russia}
% ...
\affiliation{$^{40}$Petersburg Nuclear Physics Institute,
                St. Petersburg, Russia}
% ...
\affiliation{$^{49}$Florida State University, Tallahassee, Florida 32306, USA}
% ...
\affiliation{$^{51}$University of Illinois at Chicago,
                Chicago, Illinois 60607, USA}
% ...
\affiliation{$^{59}$Kansas State University, Manhattan, Kansas 66506, USA}
% ...
\affiliation{$^{65}$Michigan State University,
                East Lansing, Michigan 48824, USA}
% ...
\affiliation{$^{75}$University of Oklahoma, Norman, Oklahoma 73019, USA}
% ...
  % input Dzero author list as of Nov 25
\date{January 8, 2009}

\begin{abstract}
A search for pair production of the lightest supersymmetric partner
of the top quark, \stopone, is performed in the lepton+jets channel
using 0.9~fb$^{-1}$ of data collected by the D0 experiment. 
Kinematic differences between \ststbar\ and the dominant top quark 
pair production background are used to separate the two processes. 
First limits from Run~II of the Fermilab Tevatron Collider for the 
scalar top quark decaying to a chargino and a $b$ quark 
($\stopone \to \charginoplus b$) are obtained for scalar top quark 
masses of 130--190~GeV and chargino masses of 90--150~GeV. 
\end{abstract}

\maketitle 

Supersymmetry~\cite{susy} introduces a superpartner for each of the 
left and the right-handed top quarks. Because of the large top quark 
mass, the mixing between those two can be substantial and lead to a 
large difference in the mass eigenvalues of the two scalar top 
(``stop") quarks. Thus, the lighter stop quark \stopone\ could 
possibly be the lightest scalar quark and within reach at the 
Fermilab Tevatron Collider. In the Minimal Supersymmetric Standard 
Model (MSSM) stop quarks are produced mainly in pairs (\ststbar) via 
the strong interaction, the same mechanism as for top quark pair
production (\ttbar)~\cite{stopproduction}. The expected 
next-to-leading-order (NLO) cross section at a center of mass 
energy of 1.96~TeV for a stop quark of mass equal to 175~GeV is 
(0.58$^{+0.16}_{-0.13}$)~pb~\cite{prospino}, while for a top quark 
of the same mass the cross section is 
(6.8$\pm$0.6)~pb~\cite{topxsec}. The stop quark pair production 
cross section strongly depends on the mass of the stop quark. 

The different possible decay modes of the stop quark result in a 
number of distinct final state signatures. The branching ratios for 
stop quark decays depend on the parameters of the model, in 
particular the masses of the supersymmetric particles involved. The 
decays to a $c$ quark and the lightest neutralino 
($\stopone \to c \neutralino$)~\cite{d0stoplimit1} and to a $b$ 
quark, a lepton, and a sneutrino 
($\stopone \to b \ell^+ \tilde{\nu}_\ell$)~\cite{d0stoplimit2} have 
already been explored at D0 in Run~II of the Tevatron. For stop 
quarks lighter than the top quark the decay channel 
$\stopone \to \charginoplus b$, with subsequent decay of the 
lightest chargino $\charginoplus$ to the lightest neutralino 
$\neutralino$ and a $W$ boson, can dominate, if kinematically 
allowed. In this Letter we assume that the branching ratio 
$B(\stopone \to \charginoplus b)=1$. This channel has been explored 
only once before by the CDF collaboration in Run~I of the Tevatron 
at $\sqrt{s}=1.8$~TeV for stop quark masses of 
100--120~GeV~\cite{cdfrun1}. With a dataset more than ten times 
larger, we obtain first limits in this channel at 
$\sqrt{s}=1.96$~TeV for stop quark masses in the range 130--190~GeV.

The \ststbar\ event signature in the studied decay channel can be 
similar to the \ttbar\ signature, making it possible for the 
\ststbar\ signal to be embedded in the \ttbar\ event sample. The 
goal of this analysis is to search for this possible hidden 
admixture. The main difference relative to \ttbar\ production is 
the additional presence of neutralinos in the event. However, this 
does not lead to significantly higher missing transverse energy 
(\met), since the neutralinos are mostly produced back-to-back. We 
consider the decay channel with one $W$ boson decaying to hadrons 
and the other one to an electron or muon and a neutrino. Scenarios 
with both on-shell and off-shell $W$ bosons provide the same 
signature. The final state consists of one high-\pt\ lepton, \met\ 
from the neutrino and the neutralinos, two jets originating from 
$b$ quarks (``$b$ jets"), and two light-quark jets. This is 
referred to as the ``lepton+jets" channel. We consider twelve mass 
points, for which the studied decay can dominate. We fix the 
neutralino mass to 50~GeV, a value close to the experimental limit 
from LEP~\cite{lep}, and we vary the stop quark mass from 130 to 
190~GeV and the chargino mass from 90 to 150~GeV to obtain the 
desired event signature. For larger neutralino masses the 
signature changes and the sensitivity of this study decreases.

The search is conducted using data collected by the D0 
detector~\cite{D0RUNII} in \ppbar\ collisions at the Fermilab 
Tevatron Collider. Triggers require an electron or muon and at least 
one jet with large transverse momentum (\pt). The dataset comprises 
an integrated luminosity of 913~pb$^{-1}$ for events containing 
electrons in the final state, and 871~pb$^{-1}$ for events with 
muons.

We select events with one isolated electron with $\pt>20$~GeV and
pseudorapidity $|\eta|<1.1$, or one isolated muon with 
$\pt>20$~GeV and $|\eta|<2.0$, and $\met>20(25)$~GeV in the 
electron (muon) channel~\cite{ttxsec}. To reject events with 
mismeasured leptons, the lepton momentum vector and the $\,\met$ 
vector are required to be separated in azimuth. In addition, we 
only accept events with $\ge$3 jets, each with $\pt>15$~GeV and 
$|\eta|<2.5$, of which the jet with largest \pt\ (``leading jet") 
has to have $\pt>40$~GeV. Events with a second isolated electron or 
muon with $\pt>15$~GeV are rejected. Details about object 
identification, jet energy corrections, and trigger requirements 
can be found in Ref.~\cite{ttxsec}. In addition, we require at least 
one $b$-tagged jet in each event, where the $b$ jets are identified 
through a neural network algorithm~\cite{btag}. 

For events with four or more jets, a kinematic fitting 
algorithm~\cite{hitfit} is used to reconstruct the objects to a 
\ttbar\ hypothesis, which is used to separate \ststbar\ from \ttbar\ 
events. The fitter minimizes a $\chi^2$ statistic within the 
constraints that both candidate $W$ boson masses are 80.4~GeV and 
that the masses of the two objects reconstructed as top quarks are 
the same. The fitter considers only the four jets of highest \pt, 
uses $b$-tagging information to minimize combinatorics, and varies 
the four-vectors of the detected objects within their resolution. 
Only events for which the fit converges (86--95\% of signal events
depending on the mass point and lepton flavor) are selected for 
further analysis.

The events are classified into four distinct subsamples, according 
to jet multiplicity (3 jets or $\ge4$ jets) and lepton flavor 
(\ejets\ or \mujets). All subsamples are used to obtain the final 
limit. 

Because of their topological similarity to the signal, \ttbar\ 
events are the most challenging background. Of the other background 
processes, production of $W$ bosons in association with jets 
($W$+jets), and multijet events, where jets are misidentified as 
isolated leptons, are most important. Far smaller contributions 
arise from $Z$+jets, single top quark, and diboson production.

Except for the multijet background, the shape of distributions in
all processes are modeled through Monte Carlo (MC) simulation. The 
\ststbar\ signal is generated by \pythia~v6.323~\cite{pythia} in 
its general MSSM mode, where the top trilinear coupling $A_t$ and 
the SU(2) gaugino mass $M_2$ are varied to set the stop quark mass 
and the chargino mass, respectively. The neutralino mass is kept at 
the same value by keeping the U(1) gaugino mass $M_1$ constant. The 
\ttbar\ background is also generated by \pythia, using a top quark 
mass of 175~GeV. The $W$+jets and $Z$+jets processes are generated 
by \alpgen~2.05~\cite{alpgen} for the matrix element calculation, 
with subsequent parton showering and hadronization generated with 
\pythia. Single top quark events are generated by 
\comphep~\cite{singletopgen} and diboson production is modeled by 
\pythia. All generated events are passed through a 
\geant-based~\cite{geant} simulation of the D0 detector and 
reconstructed using the same software as for data. To improve 
agreement between data and MC simulation, additional 
corrections~\cite{ttxsec} are applied to the simulation before 
selection.

The contribution of the multijet background for each jet 
multiplicity and lepton flavor is determined from data using a 
method which exploits the fact that this background contains jets 
that mimic leptons, whereas the other processes have a true isolated 
lepton~\cite{ttxsecb}. The normalization of the $W$+jets 
background is estimated before imposing the $b$-tagging requirement, 
by subtracting from data: (i) the estimated multijet background, and 
(ii) the \ttbar, $Z$+jets, single top, and diboson contributions as 
calculated from their next-to-leading order cross 
sections~\cite{topxsec, mcxsec}. The remaining events are assumed to 
be $W$+jets background, where we have scaled the heavy flavor
component ($Wb\bar{b}$ plus $Wc\bar{c}$) by a relative factor of 
1.17$\pm$0.18. This factor was derived on a statistically 
independent sample with two jets and at least one $b$-tag.

Table~\ref{tab_expbg} shows the numbers of expected and observed 
events after the final selection, found to be in good agreement. For 
signal events the mass points with the highest and lowest event 
yield are shown as examples.

\begin{table}
\caption{\label{tab_expbg} Expected numbers of events with total 
uncertainties and observed numbers of events after the final 
selection.}
	\begin{ruledtabular}
		\begin{tabular}{lrrrr}
		Sample & \multicolumn{2}{c}{=3 jets} & \multicolumn{2}{c}{$\ge4$ jets} \\
		           & \ejets\ & \mujets\ & \ejets\ & \mujets\ \\
		\hline
        \multicolumn{5}{l}{\bf Signal} \\
        \multicolumn{5}{l}{$m_{\stopone}$[GeV]/$m_{\chargino}$[GeV]} \\
		190/150    & 3.2$^{+0.3}_{-0.3}$ & 2.2$^{+0.2}_{-0.2}$ & 2.9$^{+0.4}_{-0.4}$ & 2.1$^{+0.3}_{-0.3}$ \\
		130/90     & 10.4$^{+1.0}_{-1.4}$ & 6.5$^{+0.6}_{-0.8}$ & 5.2$^{+0.7}_{-1.1}$ & 3.2$^{+0.6}_{-0.6}$ \\
		\hline
		\multicolumn{5}{l}{{\bf Background}} \\
		\ttbar\    &  77.6$^{+16.0}_{-15.3}$ &  58.5$^{+12.0}_{-19.6}$ & 103.0$^{+22.8}_{-22.8}$ &  84.2$^{+18.0}_{-19.2}$ \\
		$W$+jets   &  67.7$^{+25.5}_{-24.4}$ &  77.4$^{+19.1}_{-19.3}$ &  17.1$^{+12.8}_{-12.8}$ &  21.6$^{+\phantom{0}7.8}_{-\phantom{0}7.0}$ \\
		$Z$+jets   &   5.2$^{+\phantom{0}1.5}_{-\phantom{0}1.1}$ &   6.9$^{+\phantom{0}2.0}_{-\phantom{0}1.3}$ &   2.8$^{+\phantom{0}0.8}_{-\phantom{0}0.7}$ &   3.3$^{+\phantom{0}0.9}_{-\phantom{0}0.8}$ \\
		Single top &   9.3$^{+\phantom{0}1.6}_{-\phantom{0}1.5}$ &   7.5$^{+\phantom{0}1.3}_{-\phantom{0}1.2}$ &   3.1$^{+\phantom{0}0.7}_{-\phantom{0}0.7}$ &   2.5$^{+\phantom{0}0.7}_{-\phantom{0}0.6}$ \\
		Diboson    &   4.2$^{+\phantom{0}1.1}_{-\phantom{0}0.9}$ &   3.8$^{+\phantom{0}1.0}_{-\phantom{0}0.9}$ &   1.4$^{+\phantom{0}0.5}_{-\phantom{0}0.3}$ &   1.2$^{+\phantom{0}0.4}_{-\phantom{0}0.3}$ \\
		Multijet   &  22.3$\pm$4.2 &   3.0$\pm$2.4 &  10.7$\pm$2.6 &   3.3$\pm$2.7 \\
		Total      & 186.2$^{+31.4}_{-29.1}$ & 157.2$^{+22.9}_{-29.6}$ & 138.1$^{+26.8}_{-26.6}$ & 116.0$^{+20.9}_{-20.9}$ \\
		\hline
		{\bf Data} & 193\phantom{0.0} & 163\phantom{0.0} & 133\phantom{0.0} & 135\phantom{0.0} \\
		\end{tabular}
	\end{ruledtabular}
\end{table}

Because of the similarity of the \ststbar\ and \ttbar\ final 
states~\cite{thesis}, a multivariate likelihood 
discriminant~\cite{lhood} is employed to discriminate between the 
two processes. We study the kinematic differences and choose the 
variables of greatest discrimination and low correlation as input 
to the multivariate discriminant. For events with three jets, where 
the two jets besides the leading $b$-tagged jet are referred to as 
light jets $j$, the following five variables are used: (i) the 
invariant mass of the three jets, 
(ii) $K_{T}^{\rm min} = \Delta R^{\rm min}_{jj} p^{\rm min}_{T}$, 
where $\Delta R^{\rm min}_{jj}$ is the minimum 
$\Delta R$~\cite{ttxsec} separation between a pair of jets (in 
rapidity-azimuth space) and $p^{\rm min}_{T}$ is the minimum jet 
\pt\ in that pair, (iii) the smaller of the $\Delta R$ separations 
between the leading $b$-tagged jet and either the lepton or the 
vector sum of the two light jets, (iv) the \pt\ of the system of 
the two light jets, and (v) the lepton-$\,\met$ transverse 
mass~\cite{mtw}. For events with four or more jets, the following 
five variables are used: (i) the top quark mass as reconstructed by 
the kinematic fitter, (ii) the scalar sum of the \pt\ of the four 
leading jets, (iii) the invariant mass of the system of the second 
and third leading jet, excluding the leading $b$-tagged jet, 
(iv) $K_{T}^{\rm min}$, and (v) the \pt\ of the fourth leading jet.

\begin{figure*}
\subfigure{\includegraphics*[width=0.49\textwidth]{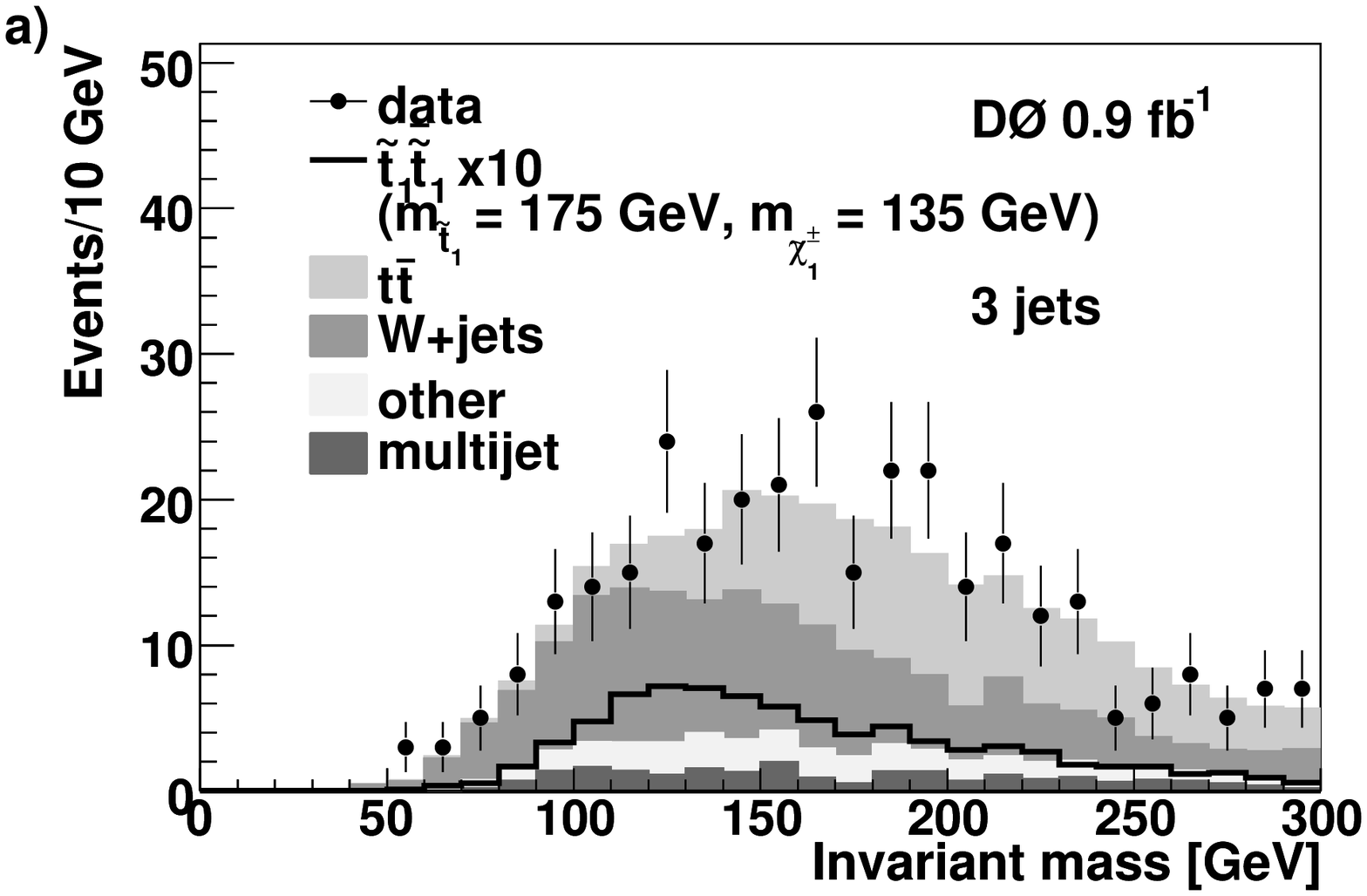}}
\subfigure{\includegraphics*[width=0.49\textwidth]{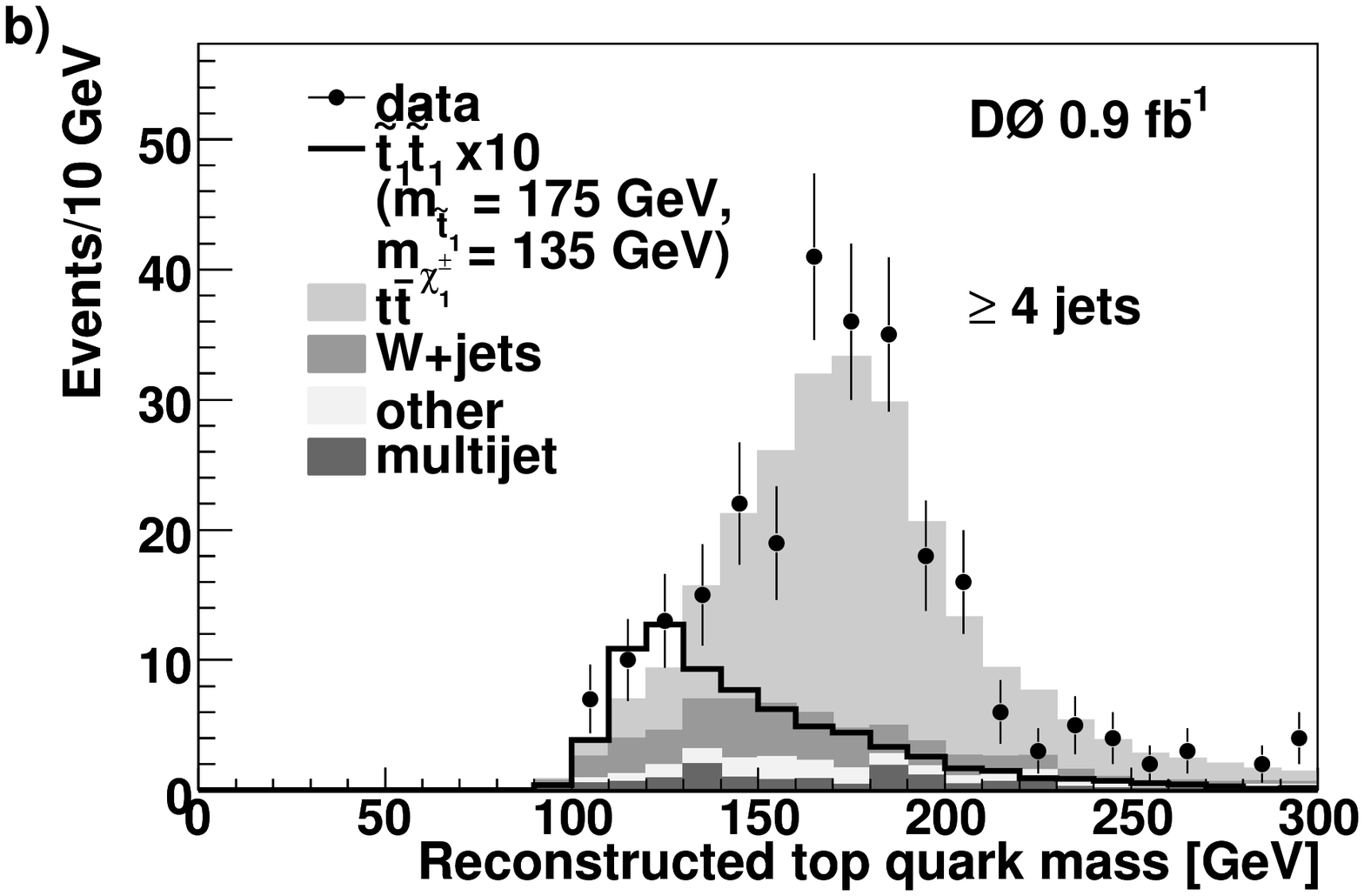}}
\caption{\label{fig_input} Comparison of the prediction with data after 
the final selection for the \ejets\ and \mujets\ channels combined, 
a) the invariant
mass of the three jets in events with 3 jets, b) the reconstructed
top quark mass in events with $\ge$4 jets. The solid line shows the 
distribution for a signal point, enhanced by a factor of ten.}
\end{figure*}

\begin{figure*}
\subfigure{\includegraphics*[width=0.49\textwidth]{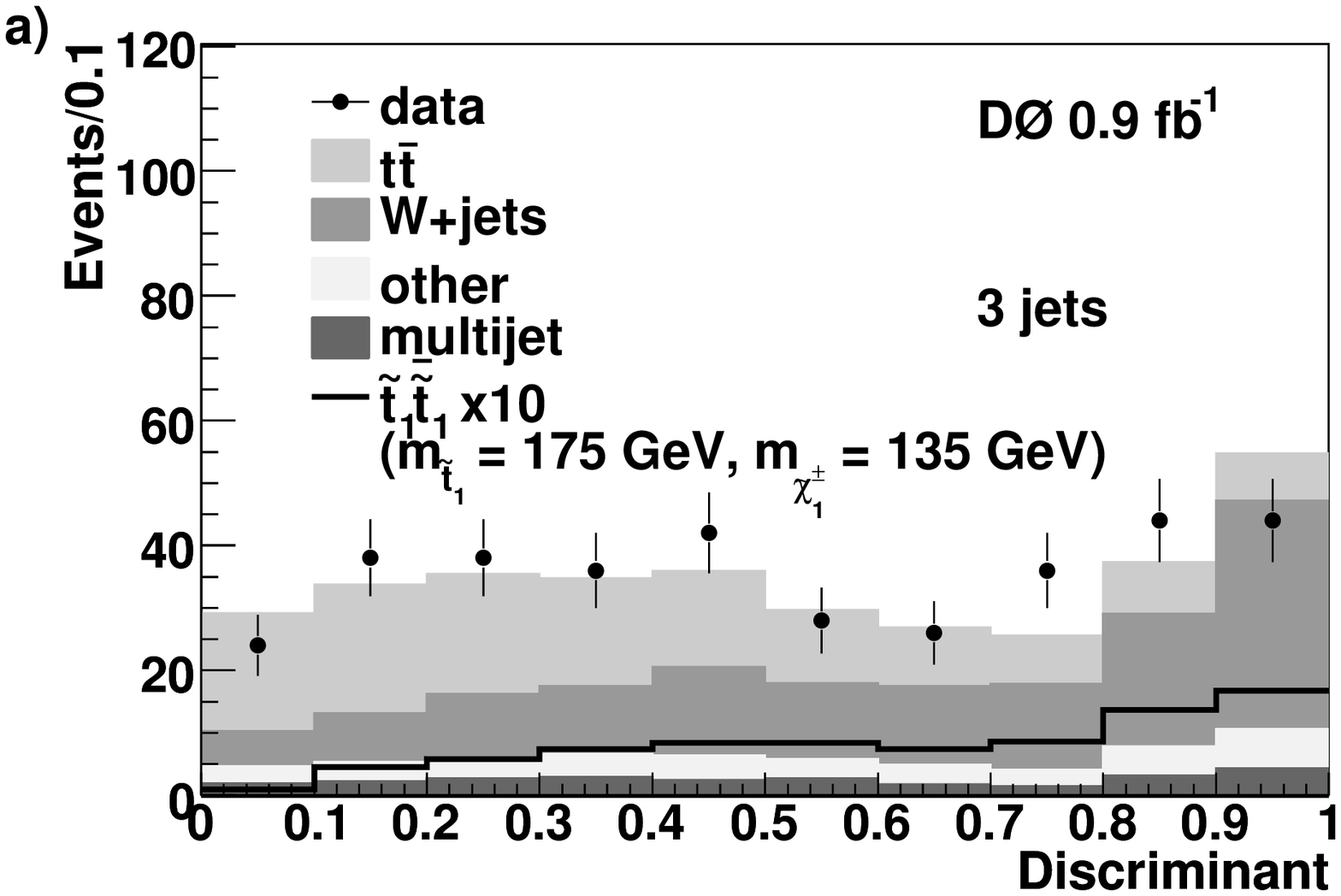}}
\subfigure{\includegraphics*[width=0.49\textwidth]{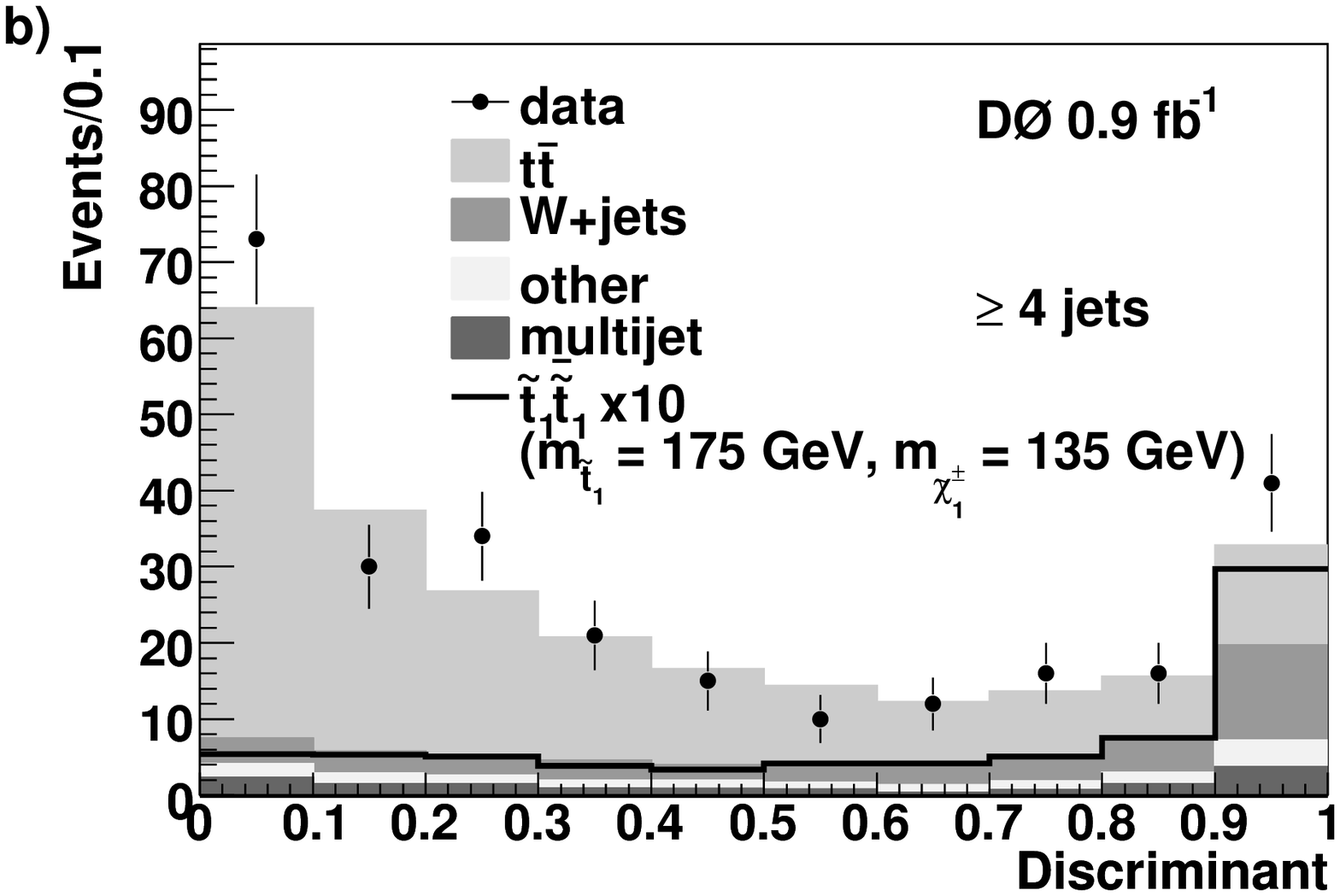}}
\caption{\label{fig_lhood} Comparison of the discriminant
distribution for data with the prediction after the final selection 
for the \ejets\ and \mujets\ channels combined, for events with 
a) 3 jets, and b) $\ge$4 jets. The solid line shows the 
distribution for a signal point, enhanced by a factor of ten.}
\end{figure*}

Figure~\ref{fig_input} shows the variable with the greatest 
separation for each jet multiplicity as a comparison between data 
and the prediction. Figure~\ref{fig_lhood} shows the resulting 
discriminant for the mass point with $m_{\stopone}$=175~GeV and 
$m_{\chargino}$=135~GeV in the 3-jet and the 4-jet subsample, 
comparing the prediction with data. The prediction for \ststbar\ 
signal (solid line) peaks at 1, while it peaks at 0 for \ttbar.

We use a Bayesian approach~\cite{limitsetting} to extract upper 
limits on the stop quark pair production cross section 
($\sigma_{\ststbar}$) from the discriminant distributions. We 
construct a binned likelihood as a product over all bins in the 
discriminant distribution as well as each of the four channels 
considered, assuming a Poisson distribution for the observed counts 
per bin. For the signal cross section, we assume a flat non-negative 
prior probability. By integrating over signal acceptance, background 
yields and integrated luminosity using Gaussian priors for each 
systematic uncertainty, we obtain the posterior probability density 
as a function of cross section for signal. The upper limit on 
$\sigma_{\ststbar}$ at 95\% confidence level is the point where the 
integral over the posterior probability density reaches 95\% of its 
total. 

We differentiate between systematic uncertainties that change the 
yield uniformly for all bins of the discriminant, and those that 
affect each bin differently. The effects are given as a percentage 
on the event yield of the affected process; they can vary widely, 
depending on the subsample and the physics process. The sources 
changing the yield uniformly include the uncertainties on integrated 
luminosity (6.1\%)~\cite{lumi}, efficiency of the event-based data 
quality selection (0.5\%), theoretical cross sections (13--20\%), 
top quark mass (1.3--7\%), estimation of the $W$+jets background 
(24--74\%, depending on the jet multiplicity and lepton flavor 
subsample), influence of the signal on the $W$+jets normalization 
(0.8--3.4\%), estimation of the multijet background (19--84\%, 
depending on the subsample), lepton identification and 
reconstruction efficiencies (2.2--2.5\%), primary vertex 
identification efficiency (2.7\%), and trigger efficiencies 
(1.2--2.7\%). The sources that also change the shape of the 
discriminant distribution include jet energy scale calibration 
(0.6--30\%), and $b$-tagging (0.1--27\%). Limits on the stop quark
pair production cross section are degraded by about a factor of two
when all systematic uncertainties are accounted for. 

\begin{figure*}
\subfigure{\includegraphics*[width=0.32\textwidth]{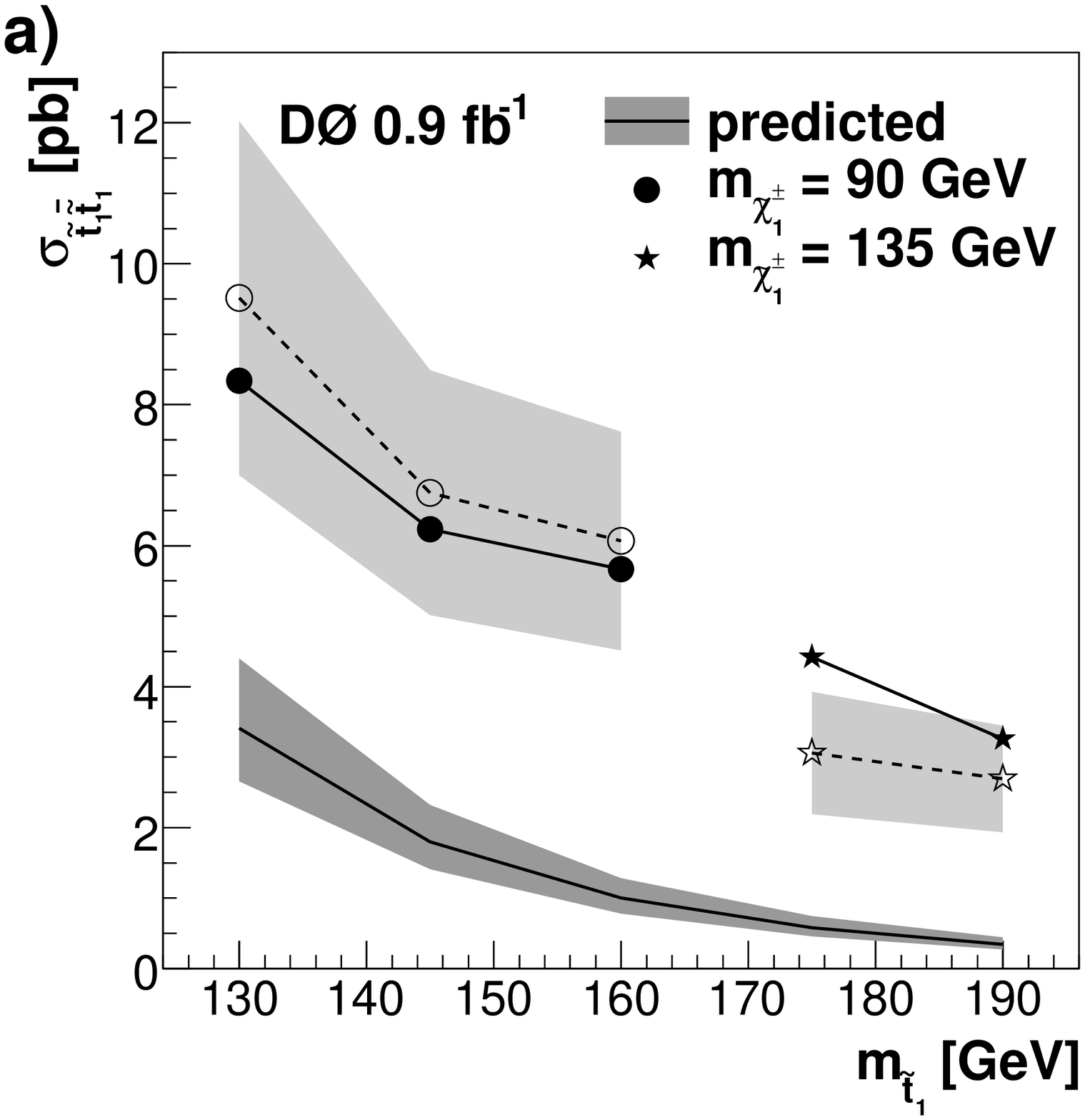}}
\subfigure{\includegraphics*[width=0.32\textwidth]{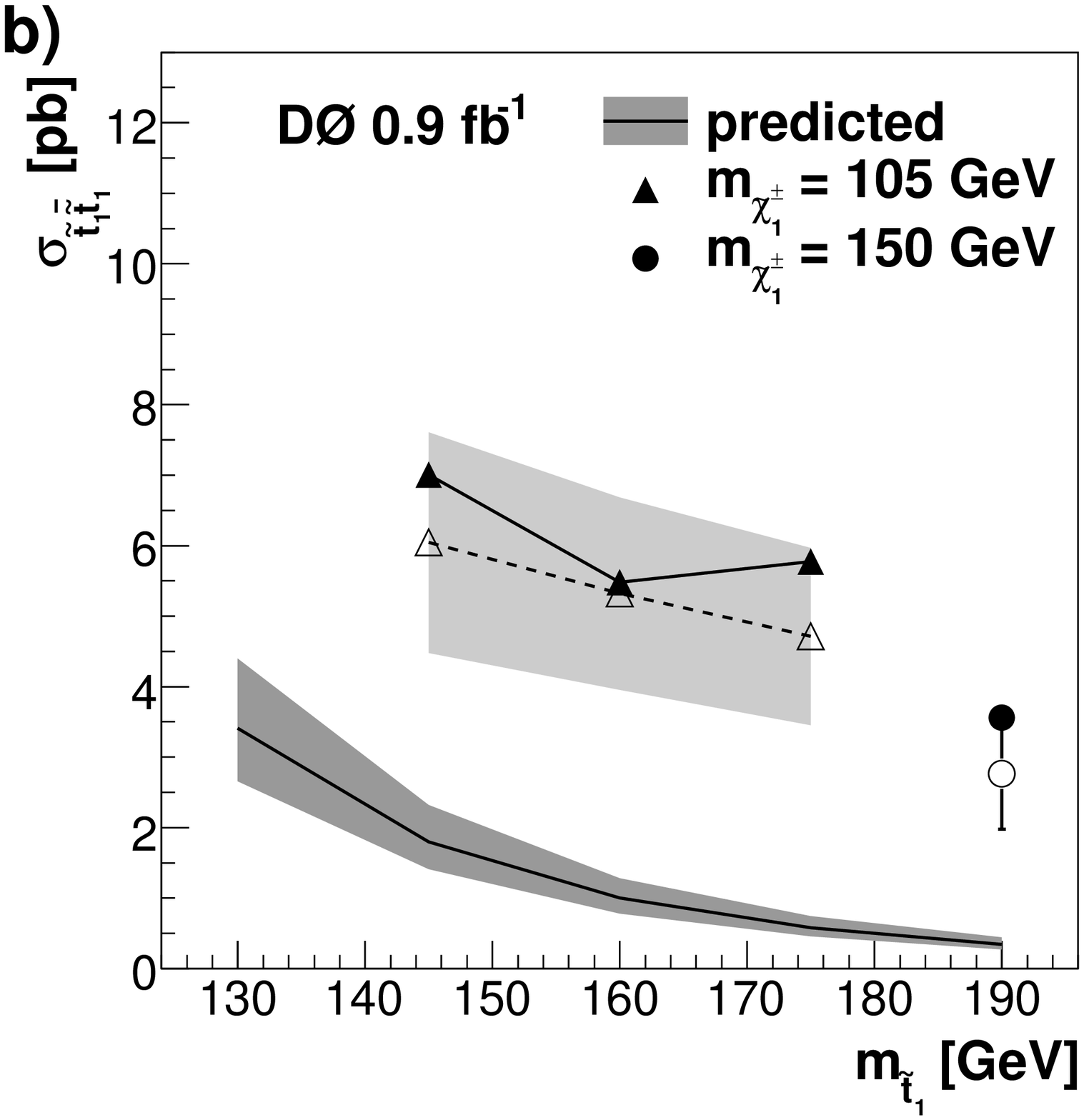}}
\subfigure{\includegraphics*[width=0.32\textwidth]{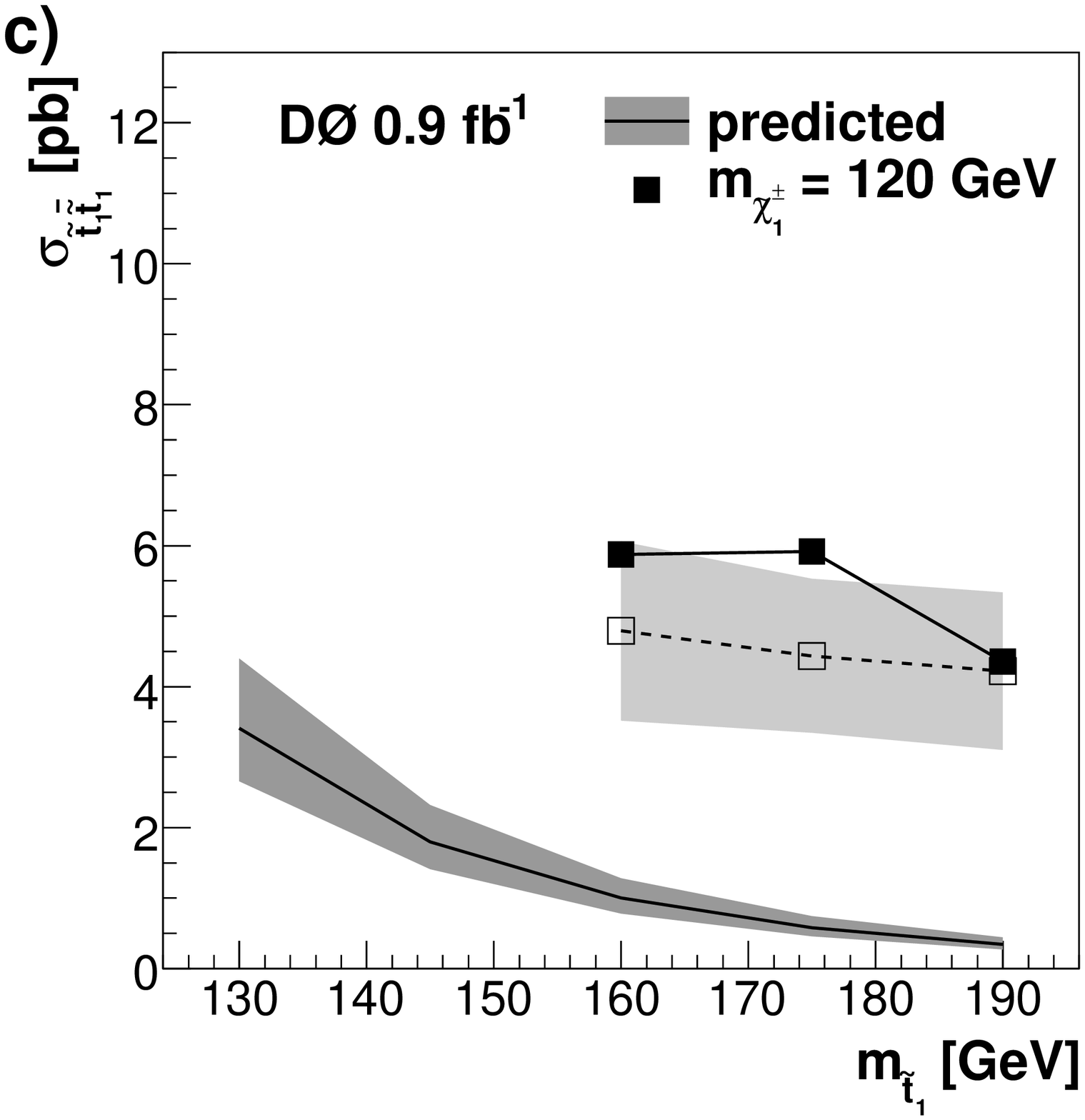}}
\caption{\label{fig_limit} Expected (open markers and dashed lines) 
and observed (filled markers and solid lines) Bayesian limits at 
95\% confidence level on the \ststbar\ cross section for all 
channels combined. Also shown is the $\pm1$ standard deviation band 
on the expected limit as well as the uncertainty on the theoretical 
prediction caused by the choice of factorization and renormalization 
scales. a) For chargino masses of 90~GeV and 135~GeV, b) for 
chargino masses of 105~GeV and 150~GeV, c) for a chargino mass of 
120~GeV.}
\end{figure*}

Table~\ref{tab_lim} shows the results for each mass point for the
combination of all channels. The results are also illustrated in 
Fig.~\ref{fig_limit}. The expected limits are derived from the sum 
of all selected background samples without a \ststbar\ contribution, 
but including the \ttbar\ background according to its theoretical 
cross section. The observed limits on the cross section are a factor 
of $2-13$ larger than the theory prediction and agree with the 
expected limits within uncertainties. In some cases, most notably 
for the mass point with $m_{\stopone}=175$~GeV and 
$m_{\chargino}=135$~GeV, the observed limit is higher than the 
expected limit, pointing to an excess of signal-like data. To 
quantify the significance, the peak position of the posterior 
probability is compared to its width. In this case, the peak is 
1.62 standard deviations away from zero. 

In summary, we present first limits on the \ststbar\ production at 
the Tevatron Run~II for a light stop quark of 130--190~GeV decaying 
to a $b$ quark and the lightest chargino. In the MSSM scenarios 
studied by this search, we derive upper limits on the cross section
that are a factor of $2-13$ above the theory prediction and agree 
with the expected limits within uncertainties.

\begin{table}
\caption{\label{tab_lim} The predicted \ststbar\ cross section and
the expected and observed Bayesian upper limits on the \ststbar\ 
cross section at the 95\% confidence level for different assumed 
values of $m_{\stopone}$ and $m_{\chargino}$. We assume 
$m_{\neutralino}=50$~GeV and $B(\stopone \to \charginoplus b)=1$.
The uncertainties on the theoretical prediction result from the 
simultaneous variation by a factor of two of the factorization and
renormalization scales about their nominal values, set equal to the 
stop quark mass. The uncertainties on the expected limits represent 
the one standard deviations estimated via background-only 
pseudo-experiments.}
	\begin{ruledtabular}
		\begin{tabular}{ccccc}
		\multicolumn{2}{c}{masses [GeV]} & \multicolumn{3}{c}{$\sigma_{\ststbar}$ [pb]} \\
		$m_{\stopone}$ & $m_{\chargino}$ & theory & exp.\ limit & obs.\ limit \\
        	\hline
		190 & 150 & 0.34$^{+0.10}_{-0.07}$ & 2.76$\pm$0.79 & 3.56 \\
		190 & 135 & 0.34$^{+0.10}_{-0.07}$ & 2.69$\pm$0.75 & 3.26 \\
		190 & 120 & 0.34$^{+0.10}_{-0.07}$ & 4.22$\pm$1.12 & 4.36 \\
		175 & 135 & 0.58$^{+0.16}_{-0.13}$ & 3.06$\pm$0.87 & 4.42 \\
		175 & 120 & 0.58$^{+0.16}_{-0.13}$ & 4.44$\pm$1.09 & 5.92 \\
		175 & 105 & 0.58$^{+0.16}_{-0.13}$ & 4.71$\pm$1.26 & 5.78 \\
		160 & 120 & 1.00$^{+0.28}_{-0.22}$  & 4.79$\pm$1.27 & 5.87 \\
		160 & 105 & 1.00$^{+0.28}_{-0.22}$  & 5.32$\pm$1.37 & 5.48 \\
		160 &  90 & 1.00$^{+0.28}_{-0.22}$  & 6.07$\pm$1.55 & 5.67 \\
		145 & 105 & 1.80$^{+0.52}_{-0.39}$  & 6.04$\pm$1.56 & 7.01 \\
		145 &  90 & 1.80$^{+0.52}_{-0.39}$  & 6.75$\pm$1.74 & 6.23 \\
		130 &  90 & 3.41$^{+0.99}_{-0.75}$  & 9.51$\pm$2.51 & 8.34 \\
        	\end{tabular}
      	\end{ruledtabular}
\end{table}

% acknowledgement_paragraph_r2.tex                         11/25/08
%
We thank the staffs at Fermilab and collaborating institutions, 
and acknowledge support from the 
DOE and NSF (USA);
CEA and CNRS/IN2P3 (France);
FASI, Rosatom and RFBR (Russia);
CNPq, FAPERJ, FAPESP and FUNDUNESP (Brazil);
DAE and DST (India);
Colciencias (Colombia);
CONACyT (Mexico);
KRF and KOSEF (Korea);
CONICET and UBACyT (Argentina);
FOM (The Netherlands);
STFC (United Kingdom);
MSMT and GACR (Czech Republic);
CRC Program, CFI, NSERC and WestGrid Project (Canada);
BMBF and DFG (Germany);
SFI (Ireland);
The Swedish Research Council (Sweden);
CAS and CNSF (China);
and the
Alexander von Humboldt Foundation (Germany).
%
   % input acknowledgement


\begin{thebibliography}{99}

  	% list_of_visitor_addresses_r2.tex                         11/25/08
%  available symbols are:
%  $\ast, \dag, \ddag, \S, \P, $\|$, $\ast\ast$, \dag\dag, \ddag\ddag ,\#
%
\bibitem[a]{alton}
Visitor from Augustana College, Sioux Falls, SD, USA.
\bibitem[b]{askew,gershtein}
Visitor from Rutgers University, Piscataway, NJ, USA.
\bibitem[c]{burdin}
Visitor from The University of Liverpool, Liverpool, UK.
\bibitem[d]{hensel,meyer,park,quadt}
Visitor from II. Physikalisches Institut, Georg-August-University,
  G{\"o}ttingen, Germany.
\bibitem[e]{luna-garcia}
Visitor from Centro de Investigacion en Computacion - IPN,
  Mexico City, Mexico.
\bibitem[f]{podesta-lerma}
Visitor from ECFM, Universidad Autonoma de Sinaloa, Culiac\'an, Mexico.
\bibitem[g]{voutilainen}
Visitor from Helsinki Institute of Physics, Helsinki, Finland.
\bibitem[h]{weber}
Visitor from Universit{\"a}t Bern, Bern, Switzerland.
\bibitem[i]{wenger}
Visitor from Universit{\"a}t Z{\"u}rich, Z{\"u}rich, Switzerland.
%\bibitem[?]{coadou}
%Visitor from Simon Fraser University, Vancouver, B.C., Canada.
%\bibitem[?]{kozminski}
%Visitor from Lewis University, Romeoville, IL, USA.
\bibitem[\ddag]{deceased}
Deceased.

%
\vskip 0.25cm
  % input visitors address
	
	\bibitem{susy}
    H.~Baer, X.~Tata,	
    Cambridge University Press (2006).	
		
	\bibitem{stopproduction} 
	W.~Beenakker et al.,
	Nucl.\ Phys.\ B515 (1998) 3.

	\bibitem{prospino} 
	W.~Beenakker et al.,
	arXiv:hep-ph/9611232v1 (1996).
	\url{http://www.ph.ed.ac.uk/~tplehn/prospino/}.
		
	\bibitem{topxsec}
	N.~Kidonakis, R.~Vogt,
  	Phys.\ Rev.\  D 68 (2003) 114014
	and private communications.

	\bibitem{d0stoplimit1}
	D\O\ Collaboration, V.~M.~Abazov et al.,
	Phys.\ Lett.\ B 665 (2008) 1.
	
	\bibitem{d0stoplimit2}
	D\O\ Collaboration, V.~M.~Abazov et al.,
	arXiv:0811.0459 [hep-ex] (2008), submitted to Phys.\ Lett.\ B.
	
	\bibitem{cdfrun1}
	CDF Collaboration, T.~Affolder et al.,
	Phys.\ Rev.\ Lett.\ 84 (2000) 5273.
	
	\bibitem{lep}
	LEPSUSYWG, ALEPH, DELPHI, L3, and OPAL Collaborations, 
	note LEPSUSYWG/01-07.1 
	(\url{http://lepsusy.web.cern.ch/lepsusy/Welcome.html}).

    	\bibitem{D0RUNII} 
    	D\O\ Collaboration, V.~M.~Abazov et al.,
    	Nucl.\ Instrum.\ Methods A 565 (2006) 463.
 
 	\bibitem{ttxsec}
	D\O\ Collaboration, V.~M.~Abazov et al.,
	Phys.\ Rev.\ D 76 (2007) 092007.

	\bibitem{btag}
	T.~Scanlon,
	Ph.D.\ Thesis, University of London, FERMILAB-THESIS-2006-43 (2006).
	
	\bibitem{hitfit}
	S.~S.~Snyder,
	Ph.D.\ Thesis, State University of New York, Stony Brook, 
	FERMILAB-THESIS-1995-27 (1995).
	
	\bibitem{pythia}
  	T.~Sj\"ostrand, L.~L\"onnblad, S.~Mrenna, P.~Skands,
   	arXiv:hep-ph/0308153 (2003).
	
	\bibitem{alpgen}
	M.~L.~Mangano et al.,
	J.\ High Energy Phys.\ 0307 (2003) 001.
	
	\bibitem{singletopgen}
	E.~E.~Boos et al.,
	Phys.\ Atom.\ Nucl.\ 69 (2006) 1317.

  	\bibitem{geant} 
 	R.~Brun, F.~Carminati, 
	CERN Program Library Long Writeup W5013, 1993 (unpublished).

	\bibitem{ttxsecb}
	D\O\ Collaboration, V.~M.~Abazov et al.,
  	Phys.\ Rev.\  D 74 (2006) 112004.
	
	\bibitem{mcxsec}
	Z.~Sullivan,
	Phys.\ Rev.\ D 70 (2004) 114012;
	J.~M.~Campbell, R.~K.~Ellis,
	Phys.\ Rev.\ D 60 (1999) 113006.
	
	\bibitem{thesis}
	S.-J.~Park,
	Ph.D.\ Thesis, University of Rochester, 
	FERMILAB-THESIS-2007-45, Appendix~B.
	
	\bibitem{lhood}
	D\O\ Collaboration, V.~M.~Abazov et al.,
	Phys.\ Lett.\ B 626 (2005) 45.
	
	\bibitem{mtw}
	J.~Smith, W.~L.~van Neerven, J.~A.~M.~Vermaseren,
	Phys.\ Rev.\ Lett.\ 50 (1983) 1738.
	
	\bibitem{limitsetting}
	D\O\ Collaboration, V.~M.~Abazov et al.,
	Phys.\ Rev.\ D 78 (2008) 012005.
	
	\bibitem{lumi}
	T.~Andeen et al.,
	FERMILAB-TM-2365 (2007).
	
\end{thebibliography}
\end{document}